\documentclass{article}
\usepackage[utf8]{inputenc}
\usepackage[width=8.5in,height=11in,margin=1in]{geometry}
\usepackage{amsmath}
\usepackage{stmaryrd}
\usepackage{graphicx}
\usepackage{listings}
\usepackage{xcolor}
\usepackage{hyperref}

\definecolor{codegreen}{rgb}{0,0.6,0}
\definecolor{codegray}{rgb}{0.5,0.5,0.5}
\definecolor{codepurple}{rgb}{0.58,0,0.82}
\definecolor{backcolour}{rgb}{0.95,0.95,0.92}

\lstdefinestyle{mystyle}{
    backgroundcolor=\color{backcolour},   
    commentstyle=\color{codegreen},
    keywordstyle=\color{magenta},
    numberstyle=\tiny\color{codegray},
    stringstyle=\color{codepurple},
    basicstyle=\ttfamily\footnotesize,
    breakatwhitespace=false,         
    breaklines=true,                 
    captionpos=b,                    
    keepspaces=true,                 
    showspaces=false,                
    showstringspaces=false,
    showtabs=false,                  
    tabsize=2
}

\title{Relational to RDF Data Migration by Query Co-Evaluation}
\author{Ryan Wisnesky, Daniel Filonik \\ Conexus AI}
\date{June 2021}

\begin{document}

\maketitle

\begin{abstract}
In this paper we define a new algorithm to convert an input relational database to an output set of RDF triples.  The algorithm can be used to e.g. load CSV data into a financial OWL ontology such as FIBO. The algorithm takes as input a set of relational conjunctive (select-from-where) queries, one for each input table; the source of each query is the three column (subject, predicate, object) RDF  schema and the target of each query is the corresponding input table's relational schema.  The algorithm’s output is the only set of RDF triples for which a unique ``round-trip'' of the input data under the relational queries exists.  The output may contain blank nodes, is unique up to unique isomorphism, and can be obtained using elementary formal methods (equational theorem proving and term model construction specifically).  We also describe how (generalized) homomorphisms between graphs can be used to write such relational conjunctive (select-from-where) queries, which, due to the lack of structure in the three-column RDF schema, tend to be large in practice.  We demonstrate examples of both the algorithm and mapping language on the FIBO financial ontology.   
\end{abstract}

\section{Introduction}

\subsection{Background}
Data migration, also known as data exchange~\cite{arenas} and closely associated with data integration~\cite{doan}, is a perennial topic in computer science, with a large body of existing results and techniques that can be immediately put to use in practical applications.  The dominant approach to both data migration and integration  (including~\cite{arenas} and~\cite{doan}) uses techniques from formal logic and model theory~\cite{abiteboul} to specify declarative ``schema mapping'' formulae relating source and target relational databases. This approach is epitomized by ETL\footnote{See section~\ref{gloss} for a glossary of acronymns.} tools such as Informatica, Altova MapForce, IBM DataStage, Ab Initio, and more, which translate the schema mappings into executable code (perhaps for a big-data engine such as Spark).  In this approach, graph data models such as RDF and  are handled by encoding graphs as their edge relations and nested data models such as XML are handled either by shredding~\cite{nested} or a higher-order logic~\cite{higher}.  Similarly, in this paper we are primarily concerned with RDF, but our technique applies any data model that can be expressed algebraically~\cite{jfp}, which includes most property graph data models.  Note that because the three-column RDF schema is a relational schema, the results in this paper apply to RDF-to-RDF translation as well, although we do not explore RDF-to-RDF translation in this paper.

\subsection{Motivation}

Although the burden of writing (or discovering) schema mappings is generally considered to be ``low'' in the literature, presumably due to their declarative character, we have found in data migration practice that schema mappings are not a substitute for queries when it come to communicating with programmers. We believe the reason for this is because query languages such as SQL are simply more widely used (and better understood) than mapping formalisms or ETL tools.  As such in this paper we propose a new method to migrate relational data to an RDF database that avoids schema mapping altogether, instead relying on a set of programmer-provided relational queries which define how to project RDF data from the relational data; these queries are then evaluated (run; executed) ``in reverse'' using a new algorithm obtained by instantiating results from a branch of mathematics known as category theory~\cite{cat}.  Although the migrated RDF data will not always be perfectly recoverable by the provided queries, the migrated data will always be ``round-trippable'' by the provided queries, allowing us to quantify the information change (gain or loss) involved in the relational to RDF translation.  We then generalize the relational to RDF migration algorithm to take into account a lightweight form of RDF schema information described in terms of (extended) graph homomorphisms, which allows us to restructure RDF databases along schema mappings in a lightweight way, as well as construct the relational  queries our algorithm requires in a less verbose way.

\subsection{Outline}
In this paper we assume familiarity with RDF~\cite{rdfsurv} and the basics of database theory~\cite{abiteboul}.  In some sections, we will occasionally make use of terminology from category theory~\cite{cat} to describe how a result is obtained, but a knowledge of category theory is not required to understand our results.  This paper is structured in two parts: first, we define our relational-to-RDF algorithm operationally, using a short example, and then describe it on a larger FIBO-based example (Section~\ref{walkthrough}).  Then, we define the schema mapping language associated to our algorithm and show how it makes the previous example more user friendly.  We have tried to avoid formal definitions in the body of the paper, but see~\cite{jfp} for the relevant mathematics.  

\subsection{For Financial Services Practitioners}

In this section we briefly discuss why our technology is ``semantic'' and the benefit its semantic aspect provides from a solution-oriented perspective.  Our technology is semantic because it ``operates up to meaning''.  That is, if two users write two syntactically different SQL queries to load data into FIBO (say they differ in numbers of FROM clauses), but those two sets of queries are semantically equivalent (always give the same result), then our technology will behave the same on both.  That is, only the meaning-- only the semantics-- of the user query matters to our technology.  Thus users of our solution need not worry about {\it how} they specify their rdf to relational transformation, only {\it what} that that transformation {\it is}.  In practice, this allows users of our technology to work more quickly, and more accurately, than users of non-semantic tools.  

\subsection{For AI Implementors}

In this section we briefly discuss why our technology is ``AI'' and the associated business cases and technical capabilities its AI aspect enables.  As can be directly seen by its use of equational reasoning and model construction, our technology is a symbolic AI, an ``expert system'' that can automatically reason in a way that respects the rules (``semantics'') of SQL queries.  Much like in machine learning (non-symbolic AI), raw computing speed enables our algorithm to operate on data sets that would have been intractable in previous decades, such as FIBO.  Relational to RDF migration itself appears as a sub-task in a wide variety of financial data analysis scenarios, for example, any use case involving both FIBO and an enterprise's non-public SQL data.

\section{Relational to RDF Migration}
\label{walkthrough}

Our algorithm migrates data from a relational database to an RDF database by evaluating a set of relational conjunctive (select-from-where) queries ``in reverse'', and then quantifying round-trip information change.  In the next sub-sections we give an operational description of our algorithm and apply it to a simple example.  The subsequent section concludes the paper by describing the associated graph-based schema mapping formalism.

\subsection{Input data preparation}
For simplicity we begin by describing the algorithm when the source relational database schema $S$ consists of exactly one table, which we for example take to be {\sf Person}, with columns named {\sf name} and {\sf age}.  In this example, we assume that the input relation for {\sf Person} contains two rows, which we will refer to as $a$ and $b$.  Our algorithm begins by taking the input table {\sf Person}, adding a {\sf row} column to it, and create new row IDs for each row; the row IDs themselves will be used later in our algorithm, and the actual values of the IDs do not matter/can be arbitrarily chosen.  That is, in this example starting from the table on the left we start by creating the table on the right:
\begin{center}
 \begin{tabular}{| c | c |} 
 \hline
{\sf name} & {\sf age} \\  
 \hline\hline
 {\sf Alice} & {\sf 20}  \\ \hline
 {\sf Bob} & {\sf 30} \\ \hline
 \end{tabular}
 \ \hspace{1in}
  \begin{tabular}{| c | | c | c |} 
 \hline
{\sf row} & {\sf name} & {\sf age} \\  
 \hline\hline
$a$ & {\sf Alice} & {\sf 20}  \\ \hline
$b$ & {\sf Bob} & {\sf 30} \\ \hline
 \end{tabular}
 \end{center}

\subsection{Query Specification}
Because our input relational database consists of one table with the source schema {\sf Person(name, age)}, we must provide a single relational conjunctive (select/from/where) query that specifies how to populate {\sf Person} from RDF triples. Assuming that the RDF database is as a three column table represented with target schema {\sf Rdf(subject,predicate,object)}, we may project out {\sf name} and {\sf age} columns as follows (where {\sf foaf} is the ``friend of a friend'' RDF schema\footnote{\url{http://www.foaf-project.org/}}:

\begin{lstlisting}[language=SQL]
CREATE VIEW Person AS
SELECT r1.object AS name, r2.object AS age
FROM Rdf AS r, Rdf AS r1, Rdf AS r2
WHERE 
  r.subject = r1.subject    AND r1.predicate = "foaf:name" AND
  r.subject = r2.subject    AND r2.predicate = "foaf:age" AND
  r.predicate = "dfs:type"  AND r.object = "foaf:person"
\end{lstlisting}
At this point, it is worth noting that the choice of query is not unique. In fact, many choices of query are possible, and as we will see, the innate information change (gain or loss or neither) of each choice can be quantified.

\subsection{Query Co-evaluation}

At a mathematical level, the source relational schema $S$ (containing a single table, {\sf Person} in this example) and target relational schema $T$ (containing a single table, {\sf Rdf} in this example) form algebraic structures called {\it categories}~\cite{cat} and the conjunctive relational queries from $T$ to $S$ is a {\it pro-functor}~\cite{jfp} which can not only be evaluated to transform $T$-databases into $S$-databases according to the usual relational (e.g. SQL) semantics, but can be ``co-evaluated'' to turn $S$-databases into $T$-databases in the ``most data quality-preserving way possible''.  An exact definition of query co-evaluation in the most general case is described in~\cite{jfp} and implemented in the open-source CQL tool\footnote{\url{http://categoricaldata.net}}; in this section, we describe its operation directly in elementary terms.  

\begin{enumerate}
    \item For every row $p$ in {\sf Person} and every {\texttt Rdf AS} $v$ statement in the query \texttt{FROM} clause
\begin{lstlisting}[language=SQL]
Rdf AS r,  Rdf AS r1,  Rdf AS r2
\end{lstlisting}
we consider the pair $(p, v)$ to uniquely identify an output RDF row.  For example, we have six output RDF tuples, which we refer to as:
$$(a, r)	\ ,	\ (a, r_1)	\ , \	(a, r_2)	\ , \ 	(b, r)	\ ,	\ (b, r_1) 	\ , \ 	(b, r_2) $$
To determine the RDF subjects, predicates, and objects of these six output rows we will examine the \texttt{WHERE}, \texttt{FROM}, and \texttt{SELECT} clauses of the input queries in turn.

\item We first examine the \texttt{WHERE} clause:
\begin{lstlisting}[language=SQL]
  r.subject = r1.subject    AND r1.predicate = "foaf:name" AND
  r.subject = r2.subject    AND r2.predicate = "foaf:age" AND
  r.predicate = "dfs:type"  AND r.object = "foaf:person"
\end{lstlisting}
and apply it the output rows enumerated above, yielding equations:
$$(a, r_0).{\sf subject} = (a, r_1).{\sf subject} \ \ \ \ 
(a, r_1).{\sf predicate} = {\sf foaf:name} \ \ \ \
(a, r_0).{\sf subject} = (a, r_2).{\sf subject} 
$$ 
$$(a, r_2).{\sf predicate} = {\sf foaf:age}  \ \ \ \ 
(a, r_0).{\sf predicate} = {\sf rdfs:type} \ \ \ \
(a, r_0).{\sf object} = {\sf foaf:person}$$ 
$$(b, r_0).{\sf subject} = (b, r_1).{\sf subject}  \ \ \ \
(b, r_1).{\sf predicate} = {\sf foaf:name} \ \ \ \
(b, r_0).{\sf subject} = (b, r_2).{\sf subject} 
$$ 
$$
(b, r_2).{\sf predicate} = {\sf foaf:age}  \ \ \ \
(b, r_0).{\sf predicate} = {\sf rdfs:type}  \ \ \ \
(b,r_0).{\sf object} = {\sf foaf:person}
$$
The above equations contain redundancy, but we have refrained from simplifying them to make it more apparent how they were computed.

\item To the equations above, we next add equations about the attributes of 
$\sf{Person}$ from the input data:
$$
a.{\sf name} = {\sf Alice} \ \ \ \ 	b.{\sf name} = {\sf Bob}	 \ \ \ \ 	a.{\sf age}={\sf 20} \ \ \ 	b.{\sf age}={\sf 30}
$$
and from \texttt{SELECT} clause of the query:

\begin{lstlisting}[language=SQL]
name AS r1.object, age AS r2.object
\end{lstlisting}
we add equations:
$$
(a, r_1).{\sf object} = a.{\sf name}	\ \ \ \ (b, r_1).{\sf object} = b.{\sf name} \ \ \ \
(a, r_2).{\sf object} = a.{\sf age}	\ \ \ \	(b, r_2).{\sf object} = b.{\sf age}
$$
Note that e.g. $a$ appears in both tuples (e.g. $(a, r_1)$) and alone (e.g. $a.{\sf name}$).

\item Finally, we determine the subjects, predicates, and objects for each output RDF row by examining the equations from the steps above and creating the (uniquely determined) minimal number of new fresh values / RDF blank nodes (indicated by ?s) required to complete the table below. We have, for example, that: 
\begin{center}
 \begin{tabular}{| c || c | c | c |} 
 \hline
 Output Row & subject & predicate & object \\  
 \hline\hline
 $(a,r)$ & ?1 & {\sf rdfs:type} & {\sf foaf:person} \\ 
 \hline
 $(b,r)$ & ?2 & {\sf rdfs:type} & {\sf foaf:person} \\
 \hline
 $(a,r_1)$ & ?1 & {\sf foaf:name} & {\sf Alice} \\
 \hline
 $(b,r_1)$ & ?2 & {\sf foaf:name} & {\sf Bob} \\
 \hline
 $(a, r_2)$ & ?1 & {\sf foaf:age} & {\sf 20} \\  
 \hline
  $(b, r_2)$ & ?2 & {\sf foaf:age} & {\sf 30} \\  
 \hline
 \end{tabular}
\end{center}
The above table can be constructed by repeatedly re-writing the equations from the above steps into a normal form, or constructing a decision procedure for the equations and then creating an initial term model, or using congruence closure algorithms, or many techniques besides.  The above RDF tuples contain two {\it blank nodes} (terms of string type not equivalent to any string)\cite{rdfsurv}, ?1 and ?2, representing Alice and Bob, respectively; the blank node $?1$ might be represented as $(a,r_0).$subject, for example. The above RDF tuples can be exported directly as e.g. XML or Turtle (.ttl), or closed under RDFS or OWL axioms as desired.  
\end{enumerate}

\subsection{Computational Complexity}

Let $r$ be the size of the input table and $q$ the size of the query.  The time complexity of the algorithm in this section is $O(rq {\sf log}(rq))$, because of the need to build a term model of a ground (variable-free) equational theory of size $rq$, such as for example with congruence closure algorithms.  In contrast, the time complexity of query evaluation is $O(r^q)$.  This analysis can be further refined to distinguish between the sizes of different parts of the query (number of joins, number of where conjuncts, etc).  Note that closing the output of our algorithm under the full OWL axioms can make computational complexity semi-computable (due to the need to decide arbitrary first-order formulas).

\subsection{Categorical Remarks}

In this section we briefly describe the categorical theory of algebraic schemas and databases, following ~\cite{jfp}. We first fix a theory, $\mathit{Ty}$, called the {\it type side} of our formalism. The sorts of $\mathit{Ty}$ are called {\it types} and the functions of $\mathit{Ty}$ are the functions that can appear in schemas and instances and queries and formulae.  The intended meaning of this theory, $\llbracket Ty \rrbracket$, is a category with products.  In this paper, our typeside has one type called {\sf Dom} (domain), and the only functions are constant symbols corresponding to strings.

A {\it schema} $S$ on type side $\mathit{Ty}$ is a theory extending $\mathit{Ty}$ with new sorts (called {\it entities}), new unary functions from entities to types (called {\it attributes}), new unary functions from entities to entities (called {\it foreign keys}), and new equations (called {\it data integrity constraints}) of the form $\forall v:s. \ t = t'$, where $s$ is an entity and $t,t'$ are terms of the same type, each containing a single free variable $v$.  The intended meaning of this theory, $\llbracket S \rrbracket$, is a category extending $\llbracket Ty \rrbracket$. The theory associated to RDF adds an entity called {\sf Rdf}, and three function symbols {\sf Rdf} $\to$ {\sf Dom} called {\sf subject}, {\sf predicate}, and {\sf object}.  The category associated to a relational schema with $N$ tables adds $N$ entities, one for each table; each column $c$ of a table $t$ adds an associated function symbol $c$ $\to$ {\sf Dom}.  

An {\it instance} $I$ on schema $S$ is a theory extending $S$ with new 0-ary function (constant) symbols called {\it generators} and non-quantified equations.  The intended meaning of an instance $I$, written $\llbracket I \rrbracket$, is the {\it term model} (i.e., {\it initial algebra}) for $I$ which contains, for each sort $s$, a {\it carrier set} consisting of the closed terms of sort $s$ modulo provability in $I$.  That is, the intended meaning of an $S$-instance $I$ is a functor $\llbracket S \rrbracket \to {\sf Set}$, namely, the initial such functor in the category of all functors consistent with $I$.  

Let $I$ and $J$ be instances on the same schema $S$.  A data mapping $h : I \Rightarrow J$ is defined as a ``derived signature morphism''~\cite{till} from $I$ to $J$ that is the identity on $\mathit{S}$.  That is, $h : I \Rightarrow J$ assigns to each generator $g : s$ in $I$ a closed term $h(g) : s$ in $J$ in a way that respects equality: if $I \vdash t = t'$, then $J \vdash h(t) = h(t')$.  A morphism of instances thus denotes a homomorphism (natural transformation) of algebras $\llbracket I \rrbracket \to \llbracket J \rrbracket$.  

Let $S$ and $T$ be schemas on the same type side $\mathit{Ty}$. A schema mapping $F : S \to T$ is defined as a ``derived signature morphism''~\cite{till} from $S$ to $T$ that is the identity on $\mathit{Ty}$.  That is, $F : S \to T$ assigns to each entity $e \in S$ an entity $F(e) \in T$, and to each attribute / foreign key $f : s \to s'$ a term $F(f)$, of type $F(s')$ and with one free variable of type $F(s)$, in a way that respects  equality: if $S \vdash t = t'$, then $T \vdash F(t) = F(t')$.  The intended meaning of $F$ is a functor $\llbracket S \rrbracket \to \llbracket T \rrbracket$ that is the identity on $Ty$.  

The database instances and morphisms on a schema $S$ constitute a category with all colimits, denoted $S{\sf-inst}$, and a schema mapping $F : S \to T$ induces a functor $\Sigma_F : S{\sf-inst} \to T{\sf-inst}$ defined by substitution.  The functor $\Sigma_F$ has a right adjoint, $\Delta_F : T{\sf-inst} \to S{\sf-inst}$, which corresponds to composition when we are thinking semantically: $\llbracket \Delta_F(I) \rrbracket = \llbracket F \rrbracket ; \llbracket I \rrbracket$.  Semantically, $\llbracket \Sigma_F(I) \rrbracket$ computes the ``left Kan-extension'' of $\llbracket I \rrbracket$ along $\llbracket F \rrbracket$~\cite{jfp}.   $\Delta_F$ has a right adjoint, $\Pi_F : S{\sf-Inst} \to T{\sf-Inst}$, which computes the right Kan-extension of $\llbracket I \rrbracket$ along $\llbracket F \rrbracket$~\cite{jfp}.

A pro-functor from category $C$ to category $D$ (written $C  \Rightarrow D$) is defined as a functor $C ^{op} \times D \to {\sf Set}$, or equivalently, by currying, $C^{op} \to {\sf Set}^D$.  A {\it query}  $Q$ from schema $S$ to schema $T$ (on the same typeside) is defined as a presentation~\cite{pro} of a pro-functor $\llbracket S \rrbracket \Rightarrow \llbracket T \rrbracket$ that maps each type to its representable instance.  $Q$ induces two adjoint data transformation operations: $ {\sf coeval}_Q  : T {\sf -Inst} \to  S {\sf -Inst}$, corresponding to composition: $ \llbracket {\sf coeval}_Q(I) \rrbracket =  \llbracket I \rrbracket ; \llbracket Q \rrbracket$ (where we implicitly convert between instances $X \to {\sf Set}$ and queries $1 \Rightarrow X$, where $1$ is the schema with a single entity); and right adjoint $  {\sf eval}_Q  :  S {\sf -Inst} \to T {\sf -Inst}$, with semantics corresponding to relational/SQL evaluation of $Q$.

In this paper we are given a select/from/where query $q_s$ (on the RDF schema) for each table $s$ in the input schema $S$, which corresponds to a query $  S   \Rightarrow  {\sf Rdf} $  as follows.  Each $q_s$ is an ${\sf Rdf}$-instance, and $ S $ has no generating morphisms between entities, so we have a functor $\llbracket S^{op} \rrbracket \to {\sf Set}^{\llbracket {\sf Rdf}\rrbracket}$; by currying, this corresponds to a functor $\llbracket S^{\sf op} \rrbracket  \times \llbracket {\sf Rdf} \rrbracket \to {\sf Set}$.  

Although ${\sf coeval}_Q$ has a direct (not necessarily `simple') categorical description (pro-functor composition), and a direct (not necessarily `simple') relational description (view unfolding), and it is even shown in~\cite{jfp} that for every $Q$, there are schema mappings $F$ and $G$ such that ${\sf coeval}_Q \cong \Sigma_F \circ \Delta_G$ (and dually for {\sf eval} and $\Pi$), co-evaluation lacks a simple algorithmic description, suitable for a wide audience, with supporting tooling for special cases of real-world practical interest, such as loading CSV data into an RDF ontology, which motivated the results in this paper.  It is our hope that anyone familiar with the basics of SQL can use the algorithm in this paper.

\subsection{Round-tripping}

The {\it round-trip} of our input data is computed by applying the user queries to the above triples, resulting in two rows on this example:

\begin{center}
 \begin{tabular}{| c || c | c |} 
 \hline
 Round-trip & {\sf name} & {\sf age}  \\  
 \hline\hline
 $r\mapsto(a,r), r_1\mapsto(a,r_1), r_2\mapsto(a,r_2)$ & Alice & 20 \\ 
 \hline
 $r\mapsto(b,r), r_1\mapsto(b,r_1), r_2\mapsto(b,r_2)$ & Bob & 30 \\
 \hline
\end{tabular}
\end{center}
Let us refer to the input relational database as $I$; then it is a theorem~\cite{jfp} that there is a unique function assigning the row IDs of the above table to the row IDs of the original source data $I$, which in this case we may write as the table:
\begin{center}
 \begin{tabular}{| c || c |} 
 \hline
 {\sf Row} & Round-trip   \\  
 \hline\hline
 $a$ & $r\mapsto(a,r), r_1\mapsto(a,r_1), r_2\mapsto(a,r_2)$  \\ 
 \hline
 $b$ & $r\mapsto(b,r), r_1\mapsto(b,r_1), r_2\mapsto(b,r_2)$  \\
 \hline
\end{tabular}
\end{center}
The above function is {\it injective} because no target row is mapped to by more than one source ID, and {\it surjective}, because every target row is mapped to by at least one source ID.  We thus say that the {\it unit}~\cite{jfp} of the relational to RDF data transformation defined in this section is {\it bijective} on the {\sf Person} table, and we conclude there is no information loss or gain.  The exact way to interpret the round-tripping function as information gain or loss is beyond the scope of this paper.  

\subsection{Multiple Input Tables}

To minimize formalism we have chosen examples that load single relational tables into RDF.  In this section, for completeness, we define the algorithm on multiple input tables, but do not provide an example.  To load a set of relational tables $R_0, \ldots, R_n$, we require a set of relational queries $Q_0 : {\sf Rdf} \to R_0, \ldots, Q_n : {\sf Rdf} \to R_n$ as usual.  In addition, for every foreign key $R_i \to R_k$, we require a ``homomorphism of queries'' from $Q_k \to Q_i$, from which it follows that ``$Q_i(J) \subseteq Q_k(J)$'' for every RDF instance $J$, as required to populate the foreign keys correctly.  We use quotes because technically we require a ``derived signature morphism'' between the ``algebraic theories'' that are induced by the queries, a notion similar in spirit but not the same as the usual notion of homomorphsim of conjunctive (select-from-where) queries, and which induces an ordering that is similar to subset containment.  In many cases, a homomorphism of queries is simply a mapping of FROM-bound variables to FROM-bound variables that preserves WHERE-clause entailment.  This additional information about how the relational queries should materialize foreign keys can be specified in the CQL tool for example, and is described in ~\cite{jfp}.

\subsection{FIBO Example}
\label{bigquery}
We conclude this section by describing a more elaborate example, of loading data into the FIBO ontology.  That is, the set of RDF triples defined in this section can be written down in .xml or .ttl form and loaded into a tool such as Protege that has been pre-populated with the FIBO ontology.   (We do note use any of the OWL axioms for FIBO; we are treating it entirely as RDF in this paper.) We begin with a single row of input payer/payee data on a single table (shown in the CQL tool\footnote{\url{http://categoricaldata.net}}):
$$
\includegraphics[width=6.5in]{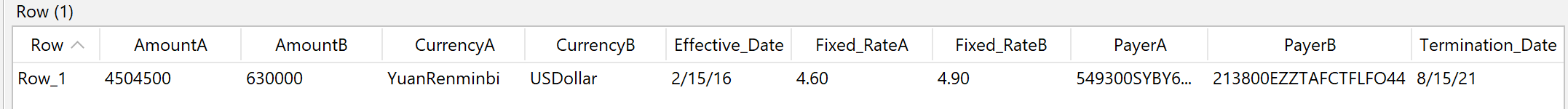} 
$$
The relational conjunctive (select-from-where) query we use to construct such a tuple from FIBO is verbose, but formulaic; later, we will see how such queries can be abbreviated.
\begin{lstlisting}[language=SQL]
SELECT    r5.object AS PayerA,            t5.object AS PayerB, 
          u3.object AS Effective_Date,    v3.object AS Termination_Date, 
          w3.object AS CurrencyA,         s3.object AS AmountA,   
          x3.object AS Fixed_RateA,       c3.object AS CurrencyB,   
          a3.object AS AmountB,           b3.object AS Fixed_RateB  
FROM R AS r1, R AS r2, R AS r3, R AS r4, R AS r5,
     R AS t1, R AS t2, R AS t3, R AS t4, R AS t5,
     R AS u2, R AS u3, R AS v2, R AS v3, R AS w2, R AS w3, R AS x2, R AS x3,
     R AS a2, R AS a3, R AS b2, R AS b3, R AS c2, R AS c3, R AS s2, R AS, s3 AS R 
WHERE							  
    r1.predicate = "...hasLeg" AND	        r1.object = r2.subject AND
    r2.predicate = "...hasPayingParty" AND 	r2.object = r3.subject AND
    r3.predicate = "...hasIdentity" AND	    r3.object = r4.subject AND
    r4.predicate = "...isIdentifiedBy" AND	r4.object = r5.subject AND
    r5.predicate = "...hasTag" AND          t1.subject = r1.subject AND
    t1.predicate = "...hasLeg" AND          t1.object = t2.subject AND
    t2.predicate = "...hasPayingParty" AND  t2.object = t3.subject AND
    t3.predicate = "...hasIdentity" AND     t3.object = t4.subject AND
    t4.predicate = "...isIdentifiedBy" AND  t4.object = t5.subject AND
    t5.predicate = "...hasTag" AND
    u2.subject = r2.subject AND             u2.predicate = "...hasEffectiveDate"  AND
    u2.object = u3.subject  AND             u3.predicate = "...hasDateValue" AND
    v2.subject = r2.subject AND             v2.predicate = "...hasTerminationDate" AND
    v2.object = v3.subject  AND             v3.predicate = "...hasDateValue" AND
    w2.subject = r2.subject AND             w2.predicate = "...hasNotationalAmount" AND
    w2.object = w3.subject  AND             w3.predicate = "...hasAmount" AND
    x2.subject = r2.subject AND             x2.predicate = "...hasInterestRate" AND
    x2.object = x3.subject  AND             x3.predicate = "...hasRateValue" AND
    a2.subject = t2.subject AND             a2.predicate = "...hasNotationalAmount" AND
    a2.object = a3.subject  AND             a3.predicate = "...hasAmount" AND
    b2.subject = t2.subject AND             b2.predicate = "...hasInterestRate" AND
    b2.object = b3.subject  AND             b3.predicate = "...hasRateValue" AND
    c2.subject = t2.subject AND             c2.predicate = "...hasCurrency" AND
    c2.object = c3.subject  AND             c3.predicate = "...hasAmount" AND
    s2.subject = r2.subject AND             s2.predicate = "...hasCurrency" AND
    s2.object = s3.subject  AND             s3.predicate = "...hasAmount"
\end{lstlisting}

When this query is co-evaluated, it creates 24 RDF triples with 15 blank RDF nodes related as shown below in the CQL tool:

$$
\includegraphics[width=6.5in]{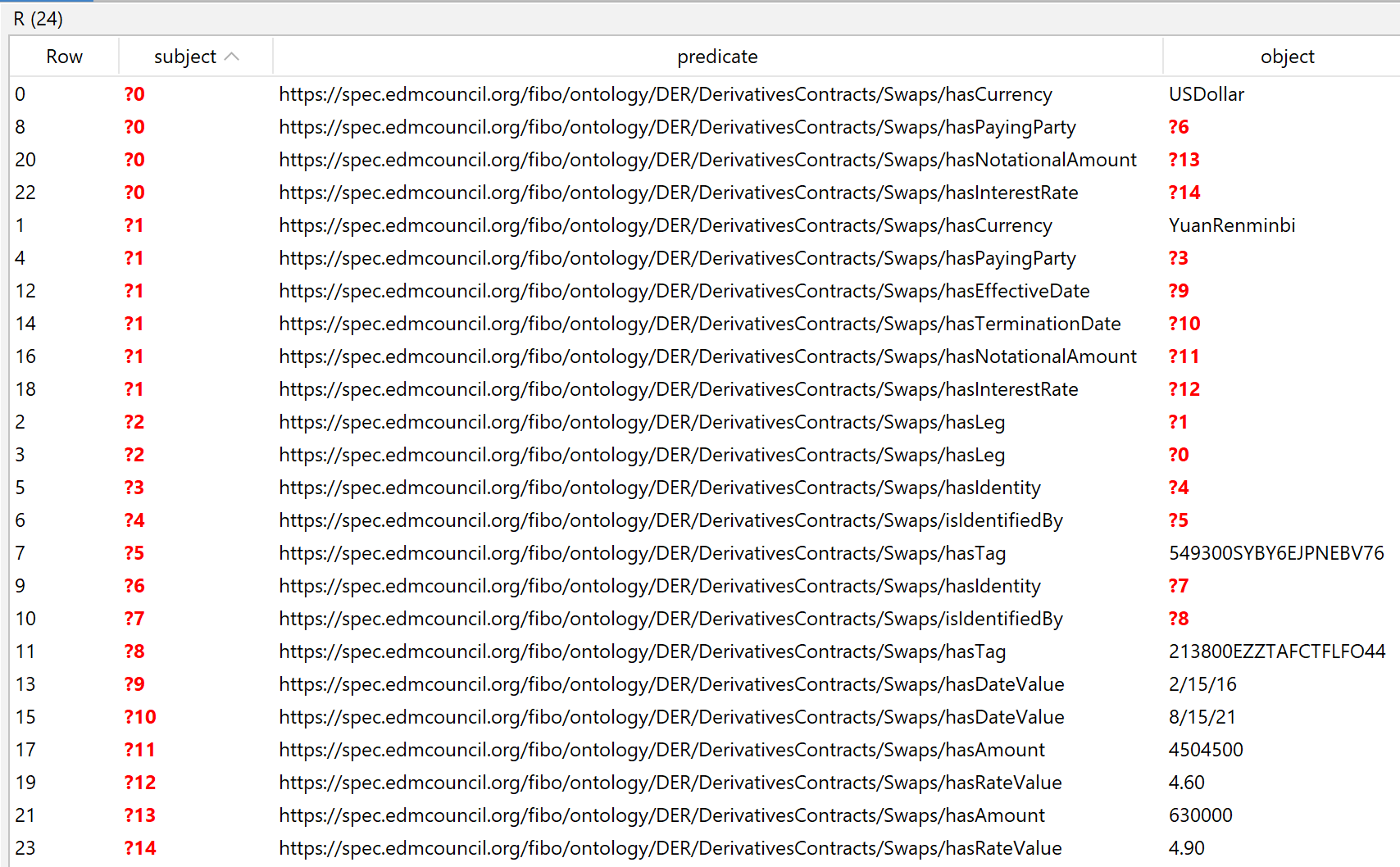}
$$

Note that every symbol, for example \texttt{USDollar}, that appears in the input row appears in the output. Note also that the blank nodes appear in multiple tuples, for example, subject ?0 has object ?13, which in turn is the subject of row 21.  

To further analyze the relational to FIBO RDF migration, we round-trip the data to find, perhaps unexpectedly, that there are two round-tripped rows:
$$
\includegraphics[width=6.5in]{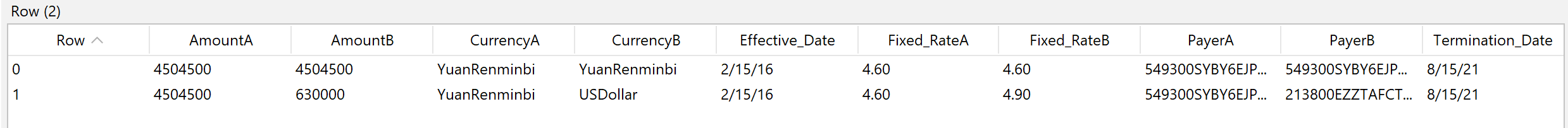}
$$
The round-trip assigns the single input row to the row with both US and Chinese currencies, but what are we to make of the additional row with two Chinese currencies, and why isn't there a row with two US currencies?  These are questions we can answer mathematically by examining the relational to RDF query.  In particular, although our query binds variables $r_1$ and $t_1$ for two ``legs'', and thus considers four possible leg-leg pairings, it only binds a single variable for an RDF triple with predicate {\sf hasEffectiveDate}, thereby restricting to just to the pairs with $r_1$ as the left component (i.e., $(t_1,r_1)$ is not considered because the query is asymmetric, at the level of syntax).  Further analysis reveals that there is no way to avoid considering all pairs of legs when we round-trip because we do not have access to the {\sf CrossCurrencyInterestRateSwap}s that describe which legs belong to the same swap.  However, even if the two legs did originate from the same swap, because in FIBO each leg has only one owner, an input row with a different buyer and seller must create two separate legs.  In other words, in FIBO, each leg has one owner, and in our input data, we have a two-legged swap; therefore, we should expect our initial row to be round-tripped into two rows.   So our conclusion is that the round-trip reveals the loading process is working exactly as intended.

As the above discussion demonstrates, the design of the relational conjunctive (select-from-where) queries input to our algorithm can have subtle implications that depend on the schema of the RDF data; the analysis in the preceding paragraph depends on the meaning of the various FIBO predicates, for example.  Fortunately, however, drawing the FIBO RDF schema as a graph can help guide the discussion, and we will see next, we can actually use a notion of graph morphism to generate the relational queries in this section.

\section{A Mapping Formalism} 

In this section we show how to formalize basic RDF schema information into graphs, and then how to translate an extended notion of graph homomorphism into the relational queries of the previous section.  

We start by defined a directed labelled multi-graph for our source data.  This graph has a node for every data type, a node for every relational table, and a directed edge for every attribute and foreign key.  Our source relational data may thus be considered as a collection of {\it sets} and {\it functions} over this graph: to each node in the graph we associate the {\it set} of rows of the associated table, and to each edge we associate a column of the input data, considered as a {\it function} from the set of rows to the domain.  Then, we define a directed labelled multi-graph for our RDF data.  This graph has a node for every data type, a node for every class, and a directed edge for every predicate.  An RDF database may thus be considered as a collection of {\it sets} and {\it relations}  over this graph: to each node in the graph we associate the set of instances of the associated class, and to each edge we associate a {\it binary relation} over the universal RDF domain.  That is, on our example we create the two graphs shown below, where the left graph for the source data has edges that mean 'function' and the right graph for the target data has edges that mean 'relation'.\footnote{Recall that a function is a special kind of relation such that if $x$ and $y$ are related, and $x$ and $y'$ are related, then $y$ must be equal to $y'$; that is, for each input, a function must give one output, whereas a relation can give zero, one, or any number of outputs to each input.  ``Father of'' is an example function, and ``is divisible by'' is an example relation.}  
$$
\includegraphics[width=5.5in]{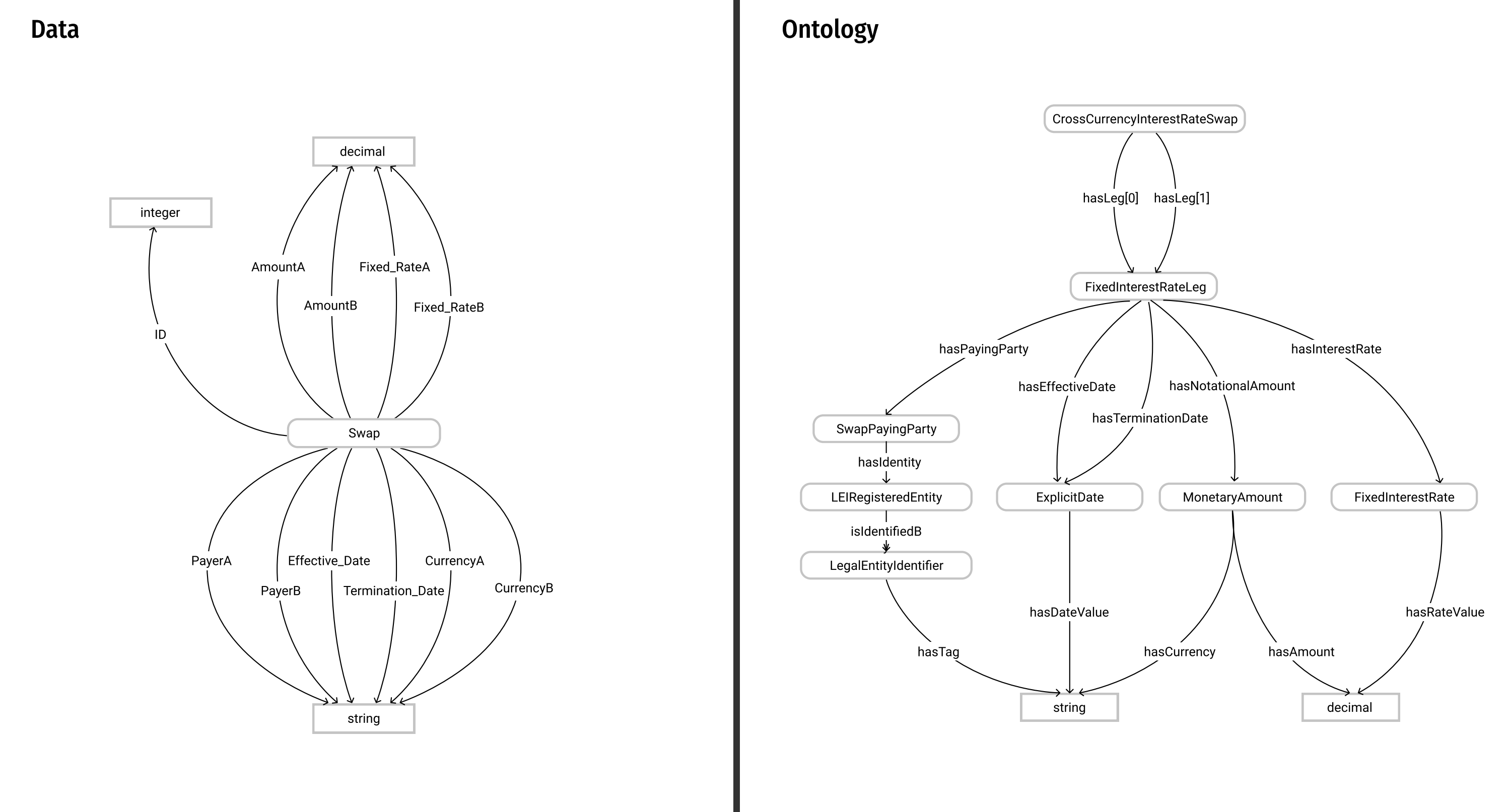}
$$
That the edges on the left graph and right graph have a different meaning is something that our algorithm mediates; for now, we define a {\it schema mapping} from the left graph to the right graph to be a function $F$ from source nodes (relational tables) to target nodes (RDF classes) and a function $F'$ from source edges (individual columns in the relational source) to paths of target edges (binary relations over the RDF domain) such that each source edge $e:s\to t$ is assigned to a path $F'(e):F(s)\to F(t)$.  This may be conveniently displayed as follows:

$$
\includegraphics[width=6.5in]{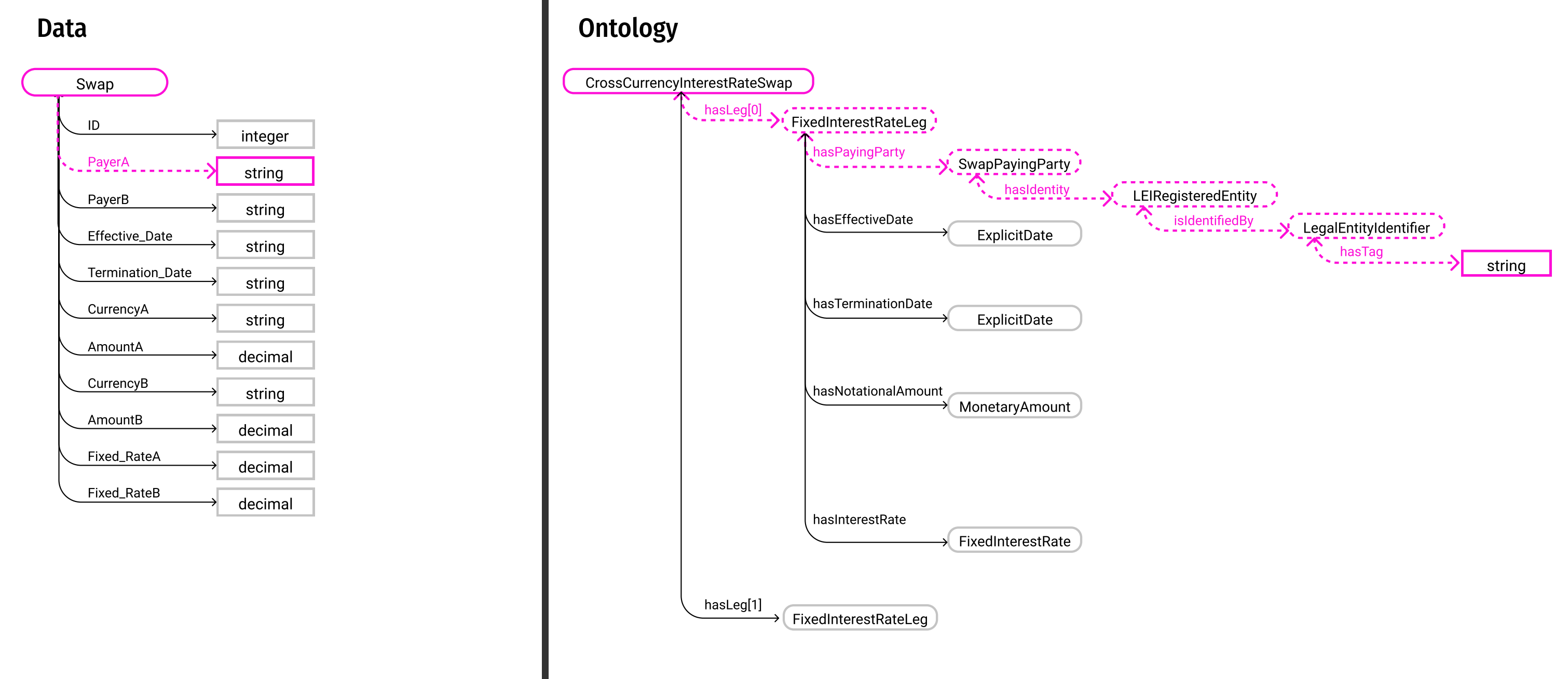}
$$

All that remains is for us to explain how to convert a schema mapping into a set of relational conjunctive (select-from-where) queries like in the previous section.  This process proceeds source table by source table.  We start by identifying the RDF class we wish to load from the table.  For example, if the source table is {\sf Swap}, we may identify the target RDF class {\sf CrossCurrencyInterestRateSwap} and begin with the following query fragment:

\begin{lstlisting}[language=SQL]
FROM 
 Rdf AS Swap 
WHERE 
 Swap.predicate = isA AND Swap.object = CrossCurrencyInterestRateSwap
\end{lstlisting}

Each column of Swap, say {\sf PayerA}, is associated with a path through the RDF schema which starts at {\sf Swap}.  Similarly to above, each edge in the RDF schema is associated with an RDF predicate, so we may consider the associated relational query, and we post-compose that query with the query we started with, repeating until we've run out of edges.  For example, after adding {\sf hasLeg[0]} to the path for {\sf PayerA} we obtain:

\begin{lstlisting}[language=SQL]
FROM 
 Rdf AS Swap, 
 Rdf AS HasLeg0 
WHERE 
 Swap.predicate = isA AND Swap.object = CrossCurrencyInterestRateSwap AND
 HasLeg0.predicate = HashLeg AND HasLeg0.subject = Swap.Object
\end{lstlisting}
Continuing this process, finally arrive at (where we project the column name we are interested in, {\sf PayerA}):
\begin{lstlisting}[language=SQL]
SELECT HasTag.object AS HasTag
FROM 
 Rdf AS Swap,
 Rdf AS HasLeg0,
 Rdf AS HasPayingParty,
 Rdf AS HasIdentity, 
 Rdf AS IsIdentifiedBy,
 RDF AS HasTag
WHERE 
 Swap.predicate = "isA" AND Swap.object = CrossCurrencyInterestRateSwap AND
 HasLeg0.predicate = "HashLeg" AND HasLeg0.subject = Swap.Object AND
 HasPayingParty.object=HasLeg0.subject AND HasPayingParty.Predicate="HasPayingParty" AND
 HasIdentity.subject=HasPayingParty.object AND HasIdentity.predicate="HasIdentity" AND
 IsIdentifiedBy.subject=HasIdentity.object AND IsIdentifiedBy.predicate="IsIdentifiedBy" AND
 HasTag.subject=IsIdentifiedBy.Object AND HasTag.predicate="HasTag"
\end{lstlisting}

Finally, we note that because each column we are mapping belongs to the same RDF class (here, {\sf Swap}), we can repeat the above process for the other columns, and then merge the resulting queries; in this example, we may add the column for {\sf HashEffectiveDate} to arrive at:
\begin{lstlisting}[language=SQL]
SELECT HasTag.subject AS HasTag, HasEffectiveDate.subject AS HasEffectiveDate 
FROM 
 Rdf AS Swap,
 Rdf AS HasLeg0,
 Rdf AS HasPayingParty, 
 Rdf AS HasIdentity, 
 Rdf AS IsIdentifiedBy, 
 Rdf AS HasTag, 
 Rdf As HasEffectiveDate
WHERE 
 Swap.predicate=isA AND Swap.object = CrossCurrencyInterestRateSwap AND
 HasLeg0.predicate=HashLeg AND HasLeg0.subject = Swap.Object AND
 HasPayingParty.object=HasLeg0.subject AND HasPayingParty.Predicate="HasPayingParty" AND 
 HasIdentity.subject=HasPayingParty.object AND HasIdentity.predicate="HasIdentity" AND
 IsIdentifiedBy.subject=HasIdentity.object AND IsIdentifiedBy.predicate="IsIdentifiedBy" AND
 HasTag.subject=IsIdentifiedBy.Object AND HasTag.predicate="HasTag" AND
 HasEffectiveDate.object=HasLeg0.subject AND HasEffectiveDate.Predicate="HasEffectiveDate"
\end{lstlisting}
Continuing on in this way, we eventually obtain a query equivalent to the one in Subsection~\ref{bigquery}.


   


\section{Conclusion and Future Work}

We have described a new algorithm, based on elementary formal methods, to translate relational databases to RDF triples by ``co-evaluating'' a set of relational conjunctive (select-from-where) queries that specify how to ``round-trip'' the relational data in ``the best way possible'', and we have shown how to quantify the information change (loss or gain) of this algorithm by examining the induced round-trips on the input data.  Because the queries required as input to this procedure can be verbose, we define a schema mapping language to abbreviate such queries and provide a graph-based representation that can be used to construct such queries through an entirely graphical user interface.  Finally, we show how both the algorithm and schema mapping language can be applied to load Cross Currency Interest Rate swaps into the FIBO RDF/OWL ontology.  Three obvious directions for future work suggest themselves; first, extending to non-conjunctive (non-select-from-where) queries, for example allowing $r_1.{\sf object} \neq t1.{\sf object}$ in where clauses.  Secondly, having SPARQL and GraphQL versions of the queries in this paper would allow its results to be transmitted to an even larger audience. Finally, the literature on RDF to/from relational data transformation is extensive~\cite{rdfsurv}, and it is likely our algorithm, albeit having novel origins in category theory, is related to other algorithms already studied in the literature.

\section{Glossary of Acronyms}
\label{gloss}
\begin{itemize}
    \item CSV: {\it comma separated values}.  A file format using the ``.csv'' name suffix, it consists of a series of newline-delimited rows, each row of which consists of comma-delimited columns.  The first line indicates the names of the columns.  
    
    \item ETL: {\it extract transform load}.  A methodology for expressing data transformation as a sequence that first extracts data from a multi-table source system, transforms it according to formalized rules in an external engine, and then loads it into a target multi-table system.
    
    \item SQL: {\it structured query language}.  A language for expressing relational queries, invented in the late 1970s and now the dominant language for data manipulation, known for its SELECT-FROM-WHERE syntax that resembles English.
    
    \item RDF: {\it resource description framework}.  A language for expressing (subject, predicate, object) triples over a controlled url-based vocabulary invented by the W3C (the standards body for the internet) in the 2000s, it can be stored in several different file formats, including XML and TTL. 
    
    \item RDFS: {\it RDF schema}.  A controlled vocabulary for expressing meta data (typing, etc) about RDF triples. 
    
    \item OWL: {\it Web Ontology Language}.  Extends RDFS to include various logics, allowing to state data integrity rules, for example that a predicate such as {\it after} has an inverse called {\it before}.  Software inference engines derive all the facts that follow from a set of RDF triples according to an OWL ontology.  
    
    \item FIBO: {\it Financial Industry Business Ontology}.  An OWL ontology that describes the business operations of the financial industry.  For example, it defines terms such as ``credit'' and ``debit'' to be opposites.
    
    \item TTL: {\it Turtle}. A language for writing RDF used in the examples above, it is much easier to read than XML.
    \item XML: {\it Extensible Markup Language}.  Techincally readable by humans but intended primarily for computers, XML provides a way to encode many kinds of information using nested trees.

\end{itemize}

\end{document}